\documentclass[12pt]{iopart}

\newcommand{\mb}[1]{\mbox{\boldmath $#1$}}

\begin{document}

\jl{6}

\title{Cosmological models with flat spatial geometry}

\author{Carlos F. Sopuerta}

\address{Relativity and Cosmology Group, \\
School of  Computer Science and Mathematics, Portsmouth University, \\ 
PO1 2EG Portsmouth, England}                            

\address{E-mail: {\tt carlos.sopuerta@port.ac.uk}}

\begin{abstract}
The imposition of symmetries or special geometric properties on 
sub\-manifolds is less restrictive than to impose them in the full 
space-time.  Starting from this idea, in this paper we study irrotational 
dust cosmological models in which the geometry of the hypersurfaces 
generated by the fluid velocity is flat, which supposes a 
relaxation of the restrictions imposed by the Cosmological
Principle.   The method of study combines covariant and tetrad methods 
that exploits the geometrical and physical properties of these models.  
This procedure will allow us to determine all the space-times within 
this class as well as to study their properties.  Some important 
consequences and applications of this study are also discussed.
\end{abstract}




~

%
%

\section{Introduction}\label{sec1}

One of the most interesting issues in modern cosmology is the
determination of the geometry of our (local) universe, an important
piece of information to understand cosmological phenomena and to interpret 
correctly observational data from them.  This question has been 
addressed in the literature many times and from many different points of 
view.  On most occasions the starting point was the Cosmological 
Principle, which assumes homogeneity and isotropy and leads to the
Friedmann-Lema\^{\i}tre-Robertson-Walker (FLRW) space-times, 
which admit a six-parameter group of isometries whose surfaces of
transitivity are spacelike hypersurfaces of constant curvature 
(see, e.g.,~\cite{HEWE}).  Hence, the
question reduces to know whether the universe corresponds to the case of
zero (flat models), or positive (closed models), or negative 
(open models) spatial curvature.  
However, we know that the Cosmological Principle is an idealization
and that some important cosmological issues, as for instance the
description of the formation of cosmic structures, require to
consider the presence of inhomogeneities in the distribution of
matter and therefore, they must be studied using inhomogeneous
cosmological models (see~\cite{KRAS} for a review).  

The main idea on which this paper is based is that we can relax the 
implications of the Cosmological Principle in the following sense: 
Instead of considering that the restrictions coming from the 
homogeneity and isotropy assumptions are placed on the space-time 
manifold, we will consider such restrictions on certain submanifolds.  
More precisely,  we will assume that the space-time is described by 
an irrotational dust solution of Einstein's equations,  which seems 
to be a good approximation to describe the universe in the 
matter-dominated era, and then we will consider that the spacelike 
hypersurfaces orthogonal to the fluid velocity admit a six-parameter 
group of motions, i.e., a maximal group of isometries.   The important 
point is that these symmetries are not necessarily symmetries of the 
full space-time, and therefore they are less restrictive. This kind of 
symmetries, those imposed on submanifold, have been usually called 
{\em intrinsic symmetries}~\cite{COLL}.
In cosmology, intrinsic symmetries have been studied
by several authors.  In particular, Collins and Szafron~\cite{COSZ}
studied them systematically in inhomogeneous cosmological models.  
Another interesting approach is the general programme put forward by 
Stephani and Wolf~\cite{STWO}, who made a systematic study of vacuum 
and perfect-fluid space-times admitting flat hypersurfaces with 
special properties of their extrinsic curvature.

As we have said before, in this paper we will study Irrotational Dust 
Models (IDMs hereafter) in which the hypersurfaces generated by the 
fluid velocity admit a maximal group of isometries, which means they
have constant curvature.  For simplicity, we will only consider here 
the case in which this group is such that the hypersurfaces are flat
(vanishing curvature).   The interest of this study is emphasized 
by recent observations~\cite{BOOM} that point to the fact that the 
spatial geometry of the universe is close to be flat.   On the other
hand, this kind of models were already considered by Collins and 
Szafron~\cite{COSZ} (Paper III).  However, in their study they neither 
find exact solutions nor finish the study of the propagation of the 
constraints.    In the present work, we have used a different point 
of view and a different procedure to study these models.  Briefly, the 
plan of this paper is the following: In section~\ref{cova}, we will
study these models from a completely covariant point of view by using 
the fluid covariant approach.  In section~\ref{basi}, we use the
results of the covariant study to choose a preferred orthonormal
basis, which will allow us to go further into this study
and to find all the possible IDMs with flat spatial geometry.
In section~\ref{disc}, we discuss the results and their main consequences, 
as well as some other related questions.

\section{Covariant approach\label{cova}}

Irrotational dust solutions of Einstein's equations are models suitable
to describe the formation of large-scale cosmic structures as well as 
the dynamics of the Universe in the matter-dominated era.   
The energy-momentum distribution is described by the fluid 
velocity $\mb{u}$ and the energy density
associated with it, $\rho$, the energy-momentum tensor being then
\begin{equation}
T_{ab} = \rho u_au_b \,, \hspace{5mm} u^au_a=-1 \,, 
\hspace{5mm} \rho > 0 \,. \label{emte}
\end{equation}
The fluid velocity is {\it irrotational} ($u_{[a}\partial_bu_{c]}=0$),
i.e., it generates orthogonal spacelike hypersurfaces, and taking
into account the conservation equations for the energy-momentum
tensor~(\ref{emte}), we deduce that it is also {\it geodesic}
($u^b\nabla_bu^a=0$).
Hence, there is locally a function $\tau(x^a)$ which is at the
same time the proper time of the matter and the label of the 
hypersurfaces orthogonal to the fluid velocity, which will be denoted
by $\Sigma(\tau)$.  In terms of $\tau$ we can write $\mb{u}$ as follows
\begin{equation}
\vec{\mb{u}} = \mb{\partial_\tau}\,,
\hspace{1cm} \mb{u} = -\mb{d\tau} \,. \label{velo}
\end{equation}
And then, using adapted coordinates $\{\tau,y^\mu\}$, where 
$y^\mu$ are comoving geodesic normal coordinates ($u^a\partial_a
y^\mu(x^b)=0$), the line element can be written in the form
\begin{equation}
ds^2 = -d\tau^2 + h_{\mu\nu}(\tau,y^\sigma)dy^\mu dy^\nu  \,.\label{line}
\end{equation}
In this paper we will study IDMs in which the hypersurfaces $\Sigma(\tau)$ 
are flat.  They are covariantly characterized by the vanishing of the 
Riemann tensor of these 3-dimensional manifolds 
\begin{equation} 
{}^3R^a{}_{bcd}=0 \,. \label{flat}
\end{equation}
This means that for each $\tau$ we can find comoving coordinates
$y^\mu$ such that the induced metric on $\Sigma(\tau)$ becomes
the flat metric, i.e., $h_{\mu\nu}|_{\Sigma(\tau)}=\delta_{\mu\nu}$.

Taking into account the properties of these models, the fluid covariant
approach to relativistic cosmology~\cite{13CF} is an adequate
tool to study them.  The variables used in this formalism
are associated with the fluid velocity $\mb{u}$, and in our case they are:
The orthogonal projector to $\mb{u}$, $h_{ab}\equiv g_{ab}+u_au_b$, which  
at the same time is the first fundamental form of the hypersurfaces 
$\Sigma(\tau)$; the non-zero kinematical quantities, the expansion 
$\Theta\equiv\nabla_au^a$ and the shear tensor $\sigma_{ab}\equiv
\nabla_{\langle a}u_{b\rangle}$\footnote{The angular brackets on indices 
denote the spatially projected, symmetric and trace-free part: 
$A_{\langle ab\rangle}\equiv[h_{(a}{}^ch_{b)}{}^d-\textstyle{1\over3}
h^{cd}h_{ab}]A_{cd}$.}, which determine the second fundamental form of the
hypersurfaces $\Sigma(\tau)$, $K_{ab}=\textstyle{1\over3}\Theta
h_{ab}+\sigma_{ab}$; the energy density, $\rho$; and the gravito-electric 
and gravito-magnetic tensors, $E_{ab}\equiv C_{acbd}u^cu^d$ and 
$H_{ab}\equiv *C_{acbd}u^cu^d$ ($E_{\langle ab\rangle}=E_{ab}$, 
$H_{\langle ab\rangle}=H_{ab}$) respectively, where $C_{abcd}$ denotes 
the Weyl tensor and $*$ the dual operation.

The equations governing these quantities can be obtained by splitting,
with respect to the fluid velocity, the Ricci identities applied to
$\mb{u}$, $(\nabla_c\nabla_d-\nabla_d\nabla_c)u^a=R^a{}_{bcd}u^b$, and
the contracted second Bianchi identities (equivalent in four dimensions
to the non-contracted ones), $\nabla^dC_{abdc}=\nabla_{[a}T_{b]c}+
\textstyle{1\over3}g_{c[a}\nabla_{b]}T^d{}_d$ (see, e.g.,
\cite{13CF}).  In order to write explicitly these equations we also
need to split the space-time covariant derivative, $\nabla_a$, into a time
derivative along the fluid worldlines,
$\dot{A}^{a\cdots}{}_{b\cdots}\equiv u^c\nabla_c A^{a\cdots}{}_{b\cdots}$,
and a spatial covariant derivative tangent to the hypersurfaces
$\Sigma(\tau)$, $D_c A^{a\cdots}{}_{b\cdots}\equiv h^a{}_e\cdots h^f{}_b
h_c{}^d\nabla_d A^{e\cdots}{}_{f\cdots}$, where $A^{a\cdots}{}_{b\cdots}$ 
is any arbitrary tensor field. Moreover, it is very useful to
introduce the spatial divergence and 
curl of an arbitrary 2-index tensor\footnote{These definitions are
analogous to those for vector fields: 
$\mbox{div}(A)\equiv D_aA^a$ and 
$\mbox{curl}A_a\equiv\varepsilon_{abc}D^bA^c\,.$} $A_{ab}$ 
(see~\cite{RMAA} for details on the notation)
\[ \mbox{div}\,(A)_a\equiv D^bA_{ab}\,,\hspace{10mm}
\mbox{curl}\, A_{ab}\equiv\varepsilon_{cd(a}D^cA_{b)}{}^d \,, \]
where $\varepsilon_{abc}\equiv\eta_{abcd}u^d$ ($\varepsilon^{abc}
\varepsilon_{def}=3! h^{[a}{}_dh^b{}_eh^{c]}{}_f$) is the volume
3-form of the hypersurfaces $\Sigma(\tau)$, and $\eta_{abcd}$
the space-time volume 4-form.  Finally,
for two arbitrary spatial symmetric tensors, $A_{ab}$ and $B_{ab}$,  
we define the commutator as 
\[ [A,B]_{ab}\equiv 2A_{[a}{}^cB_{b]c} \,, \hspace{10mm} [A,B]_a\equiv
\textstyle{1\over2}\varepsilon_{abc}[A,B]^{bc}=\varepsilon_{abc}A^b{}_d
B^{cd} \,. \]

Using all these definitions, the resulting equations can be divided into
two differentiated groups of equations. The first one
provides the evolution of our variables along the fluid worldlines,
and usually they are called the {\em evolution} equations.  In the
case of IDMs they look as follows (see, e.g., \cite{RMAA})
\begin{equation}
\dot{h}_{ab}=0\,, \label{umet}
\end{equation}
\begin{equation}
\dot{\Theta} = -{\textstyle{1\over3}}\Theta^2-\sigma^{ab}\sigma_{ab}
-{\textstyle{1\over2}}\rho\,, \label{uexp}
\end{equation}
\begin{equation}
\dot{\sigma}_{ab} = -{\textstyle{2\over3}}\Theta\sigma_{ab}-
\sigma_{c\langle a}\sigma_{b\rangle}{}^c-E_{ab} \,, \label{ushe}
\end{equation}
\begin{equation}
\dot{\rho} = -\Theta\rho \,, \label{uden} 
\end{equation}
\begin{equation}
\dot{E}_{ab} = -\Theta E_{ab}+3\sigma_{c\langle a}E_{b\rangle}{}^c
-{\textstyle{1\over2}}\rho\sigma_{ab}+\mbox{curl}\, H_{ab} \,, \label{uele}
\end{equation}
\begin{equation}
\dot{H}_{ab} = -\Theta H_{ab}+3\sigma_{c\langle a}H_{b\rangle}{}^c
-\mbox{curl}\, E_{ab} \,. \label{umag}
\end{equation}
The other group of equations are relations between spatial derivatives
of the variables, which are usually called {\em constraint} equations.
In the case of IDMs they are
(see, e.g., \cite{RMAA})
\begin{equation}
{\cal C}^1{}_a\equiv \mbox{div}\,(\sigma)_a-{\textstyle{2\over3}}
D_a\Theta =0\,, \label{cone}
\end{equation}
\begin{equation}
{\cal C}^2{}_{ab}\equiv \mbox{curl}\,\sigma_{ab}-H_{ab} =0 \,, \label{ctwo}
\end{equation}
\begin{equation} 
{\cal C}^3{}_a\equiv \mbox{div}\,(E)_a - [\sigma,H]_a-{\textstyle{1\over3}}
D_a\rho=0 \,,  \label{cthr}
\end{equation}
\begin{equation}
{\cal C}^4{}_a\equiv \mbox{div}\,(H)_a +[\sigma,E]_a =0 \,, \label{cfou}
\end{equation}
\begin{equation}
{\cal C}^5{}\equiv \rho+\textstyle{1\over2}\sigma^{ab}\sigma_{ab}
-\textstyle{1\over3}\Theta^2-\textstyle{1\over2}{}^3R =0\,, \label{cham}
\end{equation}
\begin{equation}
{\cal C}^6{}_{ab}\equiv E_{ab}-{}^3S_{ab}+\sigma_{c<a}\sigma_{b>}{}^c
-\textstyle{1\over3}\Theta\sigma_{ab}=0\,, \label{cric}
\end{equation}
where ${}^3R$ and ${}^3S_{ab}$ are the trace and trace-free parts of
the Ricci tensor, ${}^3R_{ab}$,  of the hypersurfaces $\Sigma(\tau)$.
Here, it is worth to note that the last two constraints 
(\ref{cham},\ref{cric}) come from the Gauss equations
\begin{equation}
{}^3R^a{}_{bcd}= h^a{}_eR^e{}_{fgi}h^f{}_bh^g{}_ch^i{}_d-K^a{}_c
K_{bd}+K^a{}_dK_{bc} \,, \label{gaus}
\end{equation}
which relate the Riemann tensor of the hypersurfaces $\Sigma(\tau)$,
${}^3R^a{}_{bcd}$, to that of the space-time.  Moreover, we have
taken into account that for 3-dimensional Riemannian manifolds 
the Riemann and Ricci tensors contain equivalent information,
and that the relationship between them is
\[ {}^3R^a{}_{bcd}=-2h^a{}_{be[c}{}^3R_{d]}{}^e-\textstyle{1\over2}
h^a{}_{bcd}{}^3R \hspace{5mm} 
(h_{abcd}\equiv h_{ac}h_{bd}-h_{ad}h_{bc}) \,.  \]
Then, the consequences of (\ref{flat}) and (\ref{cham},\ref{cric}) are
\begin{equation} 
{}^3 R=0 ~ \Longrightarrow ~ \rho=
\textstyle{1\over3}\Theta^2- \textstyle{1\over2}\sigma^{ab}\sigma_{ab} 
\,, \label{eden}
\end{equation}
\begin{equation} 
{}^3 S_{ab}=0 ~ \Longrightarrow ~ 
E_{ab}=\textstyle{1\over3}\Theta\sigma_{ab}-
\sigma_{\langle a}{}^c\sigma_{b\rangle c} \,. \label{gete}
\end{equation}
We can check that using these expressions for $\rho$ and $E_{ab}$ the
constraint~(\ref{cthr}) is automatically satisfied.  Furthermore, from 
equation~(\ref{gete}) it follows that the gravito-electric and shear 
tensors commute
\begin{equation} 
[\sigma,E]_{ab} \equiv \sigma_{ac}E^c{}_b-E_{ac}
\sigma^c{}_b = 0 \,. \label{seco}
\end{equation}
And from the constraint~(\ref{cfou}), we deduce that this is equivalent
to the vanishing of the spatial divergence of the gravito-magnetic
field (see, e.g., \cite{RMAA,RMLE,SMEL})
\begin{equation} 
\mbox{div}(H)_a = 0 \,. \label{divh}
\end{equation}
IDMs with a divergence-free gravito-magnetic field have been studied 
in~\cite{RMLE,SMEL}.  In these works the consistency of the new
constraints~(\ref{seco},\ref{divh}) under evolution was studied.
The conclusion was that they lead to the following new constraint
\begin{equation} 
[\sigma,\mbox{curl}H]= 0 \,, \label{copp}
\end{equation}
and that evolving repeatedly this constraint we would get an indefinite 
chain of new constraints.  However, in the case of IDMs with flat 
spatial curvature we have two additional constraints, 
namely~(\ref{eden},\ref{gete}).  Using the evolution equations 
(\ref{umet}-\ref{umag}) it can be shown that the evolution of the first 
one, i.e.~(\ref{eden}), does not lead to any new constraint, or in other 
words, it is consistent with evolution.  On the other hand, it can be 
shown that the evolution of the second one, i.e.~(\ref{gete}), leads to 
the following new constraint
\begin{equation} 
\mbox{curl}H_{ab}=0\,. \label{cugm}
\end{equation}
In order to obtain this result we have used the following property
of the shear tensor
\[ \sigma^{cd}\sigma_{c<a}\sigma_{b>d} = \textstyle{1\over2}
(\sigma^{cd}\sigma_{cd})\sigma_{ab}\,,\]
which, in fact, is satisfied by any symmetric and trace-free tensor
in 3 dimensions. As is clear, the new constraint (\ref{cugm})
implies the constraint~(\ref{copp}).  Therefore, the only constraint
that we need to analyze in our case is~(\ref{cugm}). Before to 
do that, it is worth to note that conditions (\ref{divh},\ref{cugm}) 
imply
\[ D_aH_{bc} = D_{<a}H_{bc>} \,, \]
which, following~\cite{MAES}, means that the spatial covariant
derivative of the gravito-magnetic field has only the completely
symmetric and trace-free part, which was called the {\it distortion}
of $H_{ab}$.  In~\cite{MAES}, it was argued that this part must be
non-zero if the space-time contains gravitational waves.

Following our study, we have investigated the time evolution of the
constraint~(\ref{cugm}).  To that end, it is useful to introduce
the following definitions
\[ \hat{H}_{abc}\equiv D_{<a}H_{bc>}\,, \hspace{5mm}
\hat{\sigma}_{abc}\equiv D_{<a}\sigma_{bc>}\,, \hspace{5mm}
T_a\equiv D^b\sigma_{ba}=\textstyle{2\over3}D_a\Theta\,. \]
and the decomposition in irreducible parts, with respect to the
3-dimensional rotation group, of the covariant
derivative of the shear tensor~\cite{MAES}
\[ D_a\sigma_{bc} = \hat{\sigma}_{abc}+\textstyle{3\over5}
T_{<b}h_{c>a}-\textstyle{2\over3}H_{d<b}\,\varepsilon_{c>a}{}^d
\,. \]
Then, the result from the evolution of the constraint~(\ref{cugm}) 
can be written as follows 
\begin{eqnarray}
\fl\sigma^{cd}D_c\hat{\sigma}_{dab}-\sigma_{<a}{}^cD^d
\hat{\sigma}_{b>cd}
-\textstyle{1\over2}\sigma^{cd}D_{<a}\hat{\sigma}_{b>cd}
-\textstyle{3\over2}\hat{\sigma}_{<a}{}^{cd}
\hat{\sigma}_{b>cd}+2\varepsilon_{cd<a}\hat{\sigma}_{b>}{}^{ce}
H^d{}_e \nonumber \\ \fl +3\varepsilon_{cd<a}\hat{H}_{b>}{}^{ce}
\sigma^d{}_e+5H_{<a}{}^cH_{b>c}+\textstyle{22\over5}
\varepsilon_{cd<a}{H}_{b>}{}^cT^d+\textstyle{6\over5}\left\{
\hat{\sigma}_{abc}T^c+\sigma_{ab}D_cT^c-\sigma_{c<a}D^cT_{b>}
\right.\nonumber \\
\left. \fl -\textstyle{1\over2}\sigma_{<a}{}^cD_{b>}T_c
-\textstyle{11\over20}T_{<a}T_{b>}\right\} = 0\,. \label{cogo}
\end{eqnarray}
As we can see, this expression cannot be reduced
to an identity $0=0$ by using the previous constraints, therefore it
is a new constraint.  Moreover, it only contains spatial derivatives
of the shear tensor since $T_a$ corresponds to its divergence and
$H_{ab}$ [through the constraint ${\cal C}^2$~(\ref{ctwo})] to its curl.
The next step in the covariant analysis would be to study
the subsequent evolution of this constraint.  Taking into account
that the evolution of the shear~(\ref{ushe}) can be written simply as
[using~(\ref{gete})]
\begin{equation}
\dot{\sigma}_{ab}=-\Theta\sigma_{ab} \,, \label{shes}
\end{equation}
and the form of the commutator between time and spatial derivatives
\[ \fl [\;\dot{}\;,D_a]A^{b\cdots}{}_{c\cdots} = -(\textstyle{1\over3}
\Theta\delta_a{}^d+\sigma_a{}^d)D_dA^{b\cdots}{}_{c\cdots}
+H_a{}^d\left\{\varepsilon_{de}{}^bA^{e\cdots}{}_{c\cdots}+\cdots
+\varepsilon_d{}^e{}_cA^{b\cdots}{}_{e\cdots}+\cdots\right\} \,, \]
where $A^{b\cdots}{}_{c\cdots}$ is an arbitrary tensor, we deduce that 
all the relations we would obtain from~(\ref{cogo}) would be, like the
constraints~(\ref{cugm},\ref{cogo}), of the
form $F\,[\Theta,\sigma_{ab},D_c\sigma_{ab},D_cD_d\sigma_{ab}]=0$.
However, to find these relationships would suppose a very
big amount of calculations.  Instead of this, in the next section we
will follow the study using a preferred basis.

\section{Study in the preferred basis\label{basi}}

The important point now is the fact that from equations
(\ref{gete}) and (\ref{shes}) we can deduce that there is an
orthonormal basis adapted to the fluid velocity $\mb{u}$,
$\{\mb{u},\mb{e}_\alpha\}$ ($\mb{u}\cdot \mb{e}_\alpha=0$,
$\mb{e}_\alpha\cdot\mb{e}_\beta=\delta_{\alpha\beta}$), such that
the shear and gravito-electric tensors diagonalize simultaneously
\[ \sigma_{\alpha\beta} = E_{\alpha\beta} = 0 \hspace{4mm}
\mbox{for} \hspace{4mm} \alpha\neq \beta \,, \]
and all the basis vectors are parallelly propagated along $\mb{u}$
\[ \dot{u}^a=u^b\nabla_b u^a=0\,, \hspace{10mm}
   \dot{e}_\alpha{}^a=u^b\nabla_b e_\alpha{}^a = 0 \,. \]
This result was already shown in~\cite{RMLE} for IDMs with a
divergence-free gravito-magnetic tensor, i.e.,
satisfying~(\ref{divh}).

As is clear, the shear tensor has only two independent components in
such a basis, which can be described by the following two quantities:
$\sigma_+\equiv -3\sigma_{11}/2$ and
$\sigma_-\equiv\sqrt{3}(\sigma_{22}-\sigma_{33})/2$.
From (\ref{uexp},\ref{eden}) and (\ref{shes}), the evolution
equations for the expansion $\Theta$ and these two components of the
shear can be written as follows
\[ \dot{\Theta} = -\textstyle{1\over2}(\Theta^2+\sigma^2_++
\sigma^2_-) \,, \]
\begin{equation} 
\dot{\sigma}_+=-\Theta\sigma_+ \,, \label{speq}
\end{equation}
\begin{equation}
\dot{\sigma}_-=-\Theta\sigma_- \,, \label{smeq}
\end{equation}
where from now on, $\dot{Q}\equiv\vec{\mb{u}}(Q)=
\partial_\tau Q$ for any quantity $Q$.  The important point here is the
fact that this system of evolution equations is closed, contrary to what
happens in the general case, where $\rho$ and $E_{ab}$ appear [see
equations~(\ref{uexp},\ref{ushe})].
Using adapted coordinates $\{\tau,y^\alpha\}$ [see~(\ref{line})], we can
integrate these equations and express the result in the form
\begin{equation} 
\Theta = \frac{\dot{v}}{v} \,, \hspace{5mm}
\sigma_+= \frac{{}_o\sigma_+}{v}\,,\hspace{5mm}
\sigma_-= \frac{{}_o\sigma_-}{v}\,,\hspace{5mm} 
v=\frac{3}{4}{}_o\rho(\tau-{}_o\tau)(\tau-{}_o\tau') \,,\label{solu}
\end{equation}
where the subscript ${}_o$ on a scalar means that it does not depend
on the proper time $\tau$.  In these expressions, ${}_o\sigma_+$, 
${}_o\sigma_-$, and ${}_o\tau$ are arbitrary,
${}_o\tau'$ is given by
\[ {}_o\tau' = {}_o\tau-\frac{4}{\sqrt{3}}\frac{{}_o\sigma}{{}_o\rho}\,,
\hspace{5mm}
{}_o\sigma^2\equiv\frac{1}{3}({}_o\sigma^2_++{}_o\sigma^2_-) \,, \]
and ${}_o\rho$ can be chosen arbitrarily since, as we can see from
the equation~(\ref{solu}), $v$ is defined up to a
proportionality factor depending only on $y^\alpha$.
Then, we can partially fix $v$ by choosing ${}_o\rho$ to be a constant 
(we could also choose the value).
Moreover, introducing (\ref{solu}) into (\ref{eden}) we get
\[ \rho = \frac{{}_o\rho}{v}\,,\]
so ${}_o\rho$ is directly related to the energy density.

On the other hand, the evolution equations for the triad 
$\{\mb{e}_\alpha\}$ are
\[ \dot{e}_1{}^\mu = -\textstyle{1\over3}(\Theta-2\sigma_+)e_1{}^\mu\,,\]
\[ \dot{e}_2{}^\mu = -\textstyle{1\over3}(\Theta+\sigma_++
\textstyle{\sqrt{3}}\sigma_-)e_2{}^\mu\,,\]
\[ \dot{e}_3{}^\mu = -\textstyle{1\over3}(\Theta+\sigma_+-
\textstyle{\sqrt{3}}\sigma_-)e_3{}^\mu\,,\]
which, using (\ref{solu}), can be solved to give
\begin{equation} 
e_\alpha{}^\mu = {}_oe_\alpha{}^\mu(\tau-{}_o\tau)^{-p_\alpha}
(\tau-{}_o\tau')^{-q_\alpha}\,,\label{tria}
\end{equation}
where $q_\alpha\equiv 2/3-p_\alpha$ and
\[ \fl  p_1\equiv\frac{\sqrt{3}\,{}_o\sigma-2\,{}_o\sigma_+}
{3\sqrt{3}\,{}_o\sigma}\,,\hspace{5mm}
p_2\equiv\frac{{}_o\sigma_++\sqrt{3}({}_o\sigma+{}_o\sigma_-)}
{3\sqrt{3}\,{}_o\sigma}\,,\hspace{5mm}
p_3\equiv\frac{{}_o\sigma_++\sqrt{3}({}_o\sigma-{}_o\sigma_-)}
{3\sqrt{3}\,{}_o\sigma}\,.\]
We can check that $p_\alpha$ and $q_\alpha$ satisfy the following
relationships
\begin{equation} 
\fl p_1+p_2+p_3=p^2_1+p^2_2+p^2_3=1 ~ \Longleftrightarrow ~
q_1+q_2+q_3=q^2_1+q^2_2+q^2_3=1 \,.  \label{pqre}
\end{equation}
Moreover, the quantities ${}_oe_\alpha{}^\mu$ in (\ref{tria}) are
the components of a triad defining a Riemannian
3-dimensional geometry whose (time-independent) metric tensor is
given by ${}^3_og_{\mu\nu}=\delta_{\alpha\beta}
\,{}_oe^{\alpha}{}_\mu\,{}_oe^{\beta}{}_\nu$, being 
${}_oe^{\alpha}{}_\mu$ the inverse matrix of ${}_oe_\alpha{}^\mu$.

In this situation it is worth to note that with the results that
we have obtained, or more precisely, from the expressions for
$\Theta$, $\sigma_+$, $\sigma_-$, and ${}_oe_\alpha{}^\mu$
[equations (\ref{solu},\ref{tria})], we can know the exact dependence
of any quantity on the proper time $\tau$, which means that we have
solved completely the evolution equations for the models under
consideration.
The spatial dependence of any quantity
associated with these models is provided by the knowledge of
${}_oe_\alpha{}^\mu$, ${}_o\sigma_+$, ${}_o\sigma_-$, and ${}_o\tau$.
These quantities, together with (\ref{velo}) and (\ref{tria}), determine 
completely the space-time metric.  It is clear that the equations for them 
come from Einstein's field equations, which in our case are equivalent 
to the constraints for the shear tensor~(\ref{cone}), the {\em momentum}
constraint:
\begin{equation} 
\mbox{div}(\sigma)_\alpha-
\textstyle{2\over3}\mb{e}_\alpha(\Theta)=0 \,, \label{cosh}
\end{equation}
and the equations (\ref{flat}) projected onto the triad 
$\{\mb{e}_\alpha\}$, 
\begin{equation} 
{}^3R^\alpha{}_{\beta\lambda\delta}=
\mb{e}_\lambda(\Gamma^\alpha_{\beta\delta})-
\mb{e}_\delta(\Gamma^\alpha_{\beta\lambda})+
\Gamma^\alpha_{\epsilon\lambda}\Gamma^\epsilon_{\beta\delta}-
\Gamma^\alpha_{\epsilon\delta}\Gamma^\epsilon_{\beta\lambda}-
\gamma^\epsilon_{\lambda\delta}\Gamma^\alpha_{\beta\epsilon}
=0 \,, \label{plan}
\end{equation}
where $\Gamma^\alpha_{\beta\delta}\equiv\mb{e}^\alpha\cdot
(\nabla^{}_{\mb{e}_\delta}\mb{e}_\beta)$ are the Ricci rotation
coefficients and $\gamma^\alpha_{\beta\delta}\equiv\mb{e}^\alpha\cdot
[\mb{e}_\beta,\mb{e}_\delta]$ are the commutator functions. Both
quantities are associated with the triad $\{\mb{e}_\alpha\}$ and
are related by
\begin{equation} 
\delta_{\alpha\epsilon}\Gamma^\epsilon_{\beta\lambda} =
\delta_{\alpha\epsilon}\gamma^\epsilon_{\lambda\beta}+
\delta_{\beta\epsilon}\gamma^\epsilon_{\alpha\lambda}+
\delta_{\lambda\epsilon}\gamma^\epsilon_{\alpha\beta} \,. \label{rela}
\end{equation}
From~(\ref{tria}) we get the following expression for the
commutator functions
\begin{eqnarray}
\fl \gamma^\lambda_{\alpha\beta}= \left\{
{}^{}_o\!\gamma^{\underline{\lambda}}_{\underline{\alpha\beta}}
+\left(\delta^{\underline{\lambda}}_{\underline{\alpha}}\;
{}_oe^{}_{\underline{\beta}}{}^\nu-
\delta^{\underline{\lambda}}_{\underline{\beta}}\;
{}_oe^{}_{\underline{\alpha}}{}^\nu\right) \left[
\ln(\tau-{}_o\tau)\partial_\nu p_{\underline{\lambda}}-
(\tau-{}_o\tau)^{-1}p_{\underline{\lambda}}\partial{}_o\tau
\right. \right. \nonumber \\  \left. \left. \fl
+\ln(\tau-{}_o\tau')\partial_\nu q_{\underline{\lambda}}-
(\tau-{}_o\tau')^{-1}q_{\underline{\lambda}}\partial{}_o\tau'\right]
\right\}(\tau-{}_o\tau)^{p^{}_{\underline{\lambda}}-p^{}_{\underline{\alpha}}
-p^{}_{\underline{\beta}}}\;(\tau-{}_o\tau')^{q^{}_{\underline{\lambda}}-
q^{}_{\underline{\alpha}}-q^{}_{\underline{\beta}}}\,, \label{comf}
\end{eqnarray}
where ${}^{}_o\!\gamma^\alpha_{\beta\delta}$ are the (time-independent)
commutator functions associated with the triad $\{\mb{{}_oe_\alpha}\}$, 
and underlined indices do not follow the usual index summation convention.
Now we can introduce (\ref{solu},\ref{comf}) into equations
(\ref{cosh},\ref{plan}) and from them we will have to extract
time-independent equations for the quantities 
${}_oe_\alpha{}^\mu$, ${}_o\sigma_+$,
${}_o\sigma_-$, and ${}_o\tau$.
Indeed, the three equations contained in (\ref{cosh}) turn out to be
equivalent to the following three time-independent equations
\begin{equation} 
\fl \mb{{}_oe}_1({}_o\tau+{}_o\tau'+\frac{4}{3}\frac{{}_o\sigma_+}
{{}_o\rho}) = \frac{2}{\sqrt{3}\,{}_o\rho}\left\{
(\sqrt{3}\,{}_o\sigma_++{}_o\sigma_-){}^{}_o\!\gamma^2_{12}+
(\sqrt{3}\,{}_o\sigma_+-{}_o\sigma_-){}^{}_o\!\gamma^3_{13}\right\}\,,
\label{cos1}
\end{equation}
\begin{equation} 
\fl \mb{{}_oe}_2({}_o\tau+{}_o\tau'-\frac{2}{3\,{}_o\rho}
[{}_o\sigma_++\sqrt{3}\,{}_o\sigma_-]) =\frac{2}{\sqrt{3}\,{}_o\rho}\left\{
(\sqrt{3}\,{}_o\sigma_++{}_o\sigma_-){}^{}_o\!\gamma^1_{12}-
2\,{}_o\sigma_-\,{}^{}_o\!\gamma^3_{23}\right\}\,,\label{cos2}
\end{equation}
\begin{equation} 
\fl \mb{{}_oe}_3({}_o\tau+{}_o\tau'-\frac{2}{3\,{}_o\rho}
[{}_o\sigma_+-\sqrt{3}\,{}_o\sigma_-]) =\frac{2}{\sqrt{3}\,{}_o\rho}\left\{
(\sqrt{3}\,{}_o\sigma_+-{}_o\sigma_-){}^{}_o\!\gamma^1_{13}-
2\,{}_o\sigma_-\,{}^{}_o\!\gamma^2_{23}\right\}\,,\label{cos3}
\end{equation}
On the other hand, the equations (\ref{plan}) contain six independent
relationships which, taking into account the form of the commutators
functions~(\ref{comf}) and (\ref{tria},\ref{rela}), are linear combinations
(with time-independent coefficients) of terms of the form
\begin{equation} 
\fl (\tau-{}_o\tau)^{n_1}(\tau-{}_o\tau')^{m_1}\,, \hspace{3mm}
\mbox{and} \hspace{3mm} (\tau-{}_o\tau)^{n_2}(\tau-{}_o\tau')^{m_2}
\ln^{n_3}(\tau-{}_o\tau)\ln^{m_3}(\tau-{}_o\tau')\,, \label{lico}
\end{equation}
where the exponents $n_\alpha$, $m_\alpha$ are rational numbers which
take only some values, for instance: $n_3,m_3=0,1,2$ and $n_3+m_3=2$.
In general, from each equation in (\ref{plan}) we will
extract several relations for the quantities ${}_oe_\alpha{}^\mu$, 
${}_o\sigma_+$, ${}_o\sigma_-$, and ${}_o\tau$.  To that end, we have
carried out a careful study of the linear combinations of terms~(\ref{lico})
coming from (\ref{plan}), considering the linear interdependence 
between the terms~(\ref{lico}) and taking into account that these 
relations must be satisfied for all the fluid elements (labelled
by {\em comoving} coordinates $y^\alpha$) and for all the values
of the proper time $\tau$.

To put into practice the plan just outlined, it is very
convenient to consider two separate cases depending on whether the shear
tensor is degenerate or not, or equivalently through (\ref{gete}),
on whether the gravito-electric tensor is degenerate or not.
The shear tensor is degenerate if and only if one of the following
quantities vanishes
\[   \frac{\sqrt{3}\,{}_o\sigma_++{}_o\sigma_-}{3\,{}_o\sigma}
=p_2-p_1\,, \hspace{3mm}\frac{\sqrt{3}\,{}_o\sigma_+-{}_o\sigma_-}
{3\,{}_o\sigma}=p_3-p_1\,, 
\hspace{3mm}\frac{2\,{}_o\sigma_-}{3\,{}_o\sigma}=p_2-p_3\,.\]
That is to say, if and only if two $p_\alpha$ (or also two $q_\alpha$)
are equal.  We have analyzed both cases arriving at the following results:

\subsection{Algebraically general case}

In this case it is clear that all the $p_\alpha$ must be different. In fact,
from~(\ref{pqre}) we can assume, without loss of generality, that in an 
open domain of the space-time we have: $p_3\in (-\textstyle{1\over3},0)$,
$p_2\in (0,\textstyle{2\over3})$, and $p_1\in
(\textstyle{2\over3},1)$.  Studying the equations
(\ref{plan}) and (\ref{cos1}-\ref{cos3}) we have found that they
imply:
\[ \gamma^\alpha_{\beta\delta}=0\,, \hspace{5mm}
p_\alpha=\mbox{constant}\,, \hspace{5mm} {}_o\tau=\mbox{constant}
\,.\]
Therefore, we conclude that the only possible metrics in this case
are those belonging to the family of the Bianchi I IDMs~\cite{BIAN},
whose line element can be written as
\begin{equation} 
\mbox{ds}^2 = -d\tau^2 + \sum_{\alpha=1}^3 
\tau^{2p_\alpha}(\tau-{}_o\tau)^{2(2/3-p_\alpha)}(dy^\alpha)^2 \,.
\label{biai} 
\end{equation}

\subsection{Degenerate case}

In this case two of the $p_\alpha$ must be equal, we can choose
without loosing generality
\begin{equation} 
p_2-p_3=0 ~ \Longleftrightarrow ~ {}_o\sigma_-=0\,,\label{dege}
\end{equation}
and from (\ref{pqre}) we can take $p_1=-1/3$, $p_2=p_3=2/3$ 
($q_1=1$, $q_2=q_3=0$).  As a consequence, in this case there are fewer 
linearly-independent terms in~(\ref{lico}) and then, we get also fewer 
restrictions.  Moreover, we must take
into account that the triad $\{\mb{{}_oe}_\alpha\}$ (and also the triad
$\{\mb{e}_\alpha\}$) is now not completely fixed [due to (\ref{dege})], 
there is the freedom of a rotation on the 2-planes spanned by
$\mb{e}_2$ and $\mb{e}_3$.  Taking all these considerations into account,
we have found that the analysis of the equations
(\ref{cos1}-\ref{cos3}) and (\ref{plan}) implies that the only possible
metrics are those belonging
to the Szekeres space-times~\cite{SZEK} and such that the hypersurfaces 
orthogonal to the fluid velocity are flat, which were given in~\cite{BOSP}.  
There are two families of solutions which in our formalism are 
distinguished according to whether $\mb{e}_1({}_o\tau)$
vanishes or not (in the formulation of \cite{SZEK} the key quantity is 
$\beta'\equiv\partial_r\beta$).

When $\mb{e}_1({}_o\tau)=0$, the line element is given by [$(y^\alpha)=
(x,y,z)$]
\begin{equation} 
  \mbox{ds}^2 = -d\tau^2 + \left[ (1+A y+B z)\tau+C\right]^2
\tau^{-\textstyle{2\over3}}dx^2 + \tau^{\textstyle{4\over3}}
\left[dy^2 + dz^2\right] \,, \label{cla1}
\end{equation}
where $A$, $B$ and $C$ are arbitrary functions of $x$. And when 
$\mb{e}_1({}_o\tau)\neq 0$, it can be cast into the form
\begin{equation} 
\fl \mbox{ds}^2 = -d\tau^2 + \frac{V^2[\partial_x(\ln U)]^2}
{(\tau-{}_o\tau)^{\textstyle{2\over3}}}\left[\tau-{}_o\tau + \frac{2}{3}
\frac{\partial_x({}_o\tau)}{\partial_x(\ln U)}\right]^2 dx^2 
+ U^2(\tau-{}_o\tau)^{\textstyle{4\over3}}
\left[dy^2+ dz^2\right] \,, \label{cla2}
\end{equation}
where 
\[ U = V W\,, \hspace{5mm} 
W=\left\{ a\left[y^2+z^2\right]+2b y+2c z+d\right\}^{-1} \,,\]
and where $a$, $b$, $c$, and $d$ are any functions of $x$ such that 
$ad-b^2-c^2=1$, and $V$ and ${}_o\tau$ are arbitrary functions of $x$.

\section{Analysis and discussion\label{disc}}

Now, we are going to analyze the consequences of the study that we have 
carried out in the previous sections.
First of all,  we have determined all the IDMs in which the hypersurfaces
orthogonal to the fluid velocity are flat.  More specifically, we have
shown the following statement: {\em ``All the irrotational dust solutions
of Einstein's equations with flat spatial geometry are given by the 
family of the Bianchi I dust solutions~(\ref{biai}), and by the two
subfamilies of the Szekeres dust models given 
in~(\ref{cla1},\ref{cla2})".}
Another consequence of this study is that it provides an intrinsic
space-time characterization of these solutions.  This characterization
is based on the fact that within this class of models we can distinguish
two classes according to the algebraic type of the shear tensor,
or equivalently [through the relation~(\ref{gete})], according to the
algebraic type of the gravito-electric tensor, $E_{ab}$.  But taking into
account the well-known fact that these
solutions~(\ref{biai},\ref{cla1},\ref{cla2}) have a vanishing
gravito-magnetic tensor, $H_{ab}=0$, these two classes can
be distinguished according to the Petrov type.  Then, the intrinsic
characterization is as follows:  The Bianchi I dust models~(\ref{biai})
are the {\em Petrov type I IDMs with flat spatial geometry} and the two
subfamilies~(\ref{cla1},\ref{cla2}) of the Szekeres models are
the {\em Petrov type D IDMs with flat spatial geometry}.
Both classes contain the Petrov type O models, which correspond to
the Einstein-de Sitter space-time~\cite{EIDS}.

With regard to the Killing symmetries, in the algebraically general
case the Bianchi I models are homogeneous cosmological models
in which the 3-dimensional simply-transitive group of motions is
Abelian.  In the degenerate case the situation is more interesting
since the models in this class belong to the Szekeres class of 
inhomogeneous solutions, and in~\cite{BOST} it was shown that the 
Szekeres solutions have no Killing vectors in general.  We have carried 
out the same study for the two particular families given by 
(\ref{cla1},\ref{cla2}),  and we have found that both classes of solutions 
have also in general no Killing vectors.   This results reinforces the 
idea expressed by Collins~\cite{COLL} that {\em intrinsic symmetries} 
(symmetries of submanifolds) are considerably less restrictive than 
space-time symmetries.  In our case the hypersurfaces $\Sigma(\tau)$ are 
flat, i.e., they have 6 (intrinsic) Killings, but we have found models 
without space-time Killing symmetries.

On the other hand, as we have mention before, all these
solutions~(\ref{biai},\ref{cla1},\ref{cla2}) have a vanishing 
gravito-magnetic tensor, $H_{ab}=0$.  Therefore, they belong to the class
of the {\em silent} universes~\cite{SILD},  IDMs with a vanishing 
gravito-magnetic tensor.  These models have attracted much attention 
in the last years in the context of structure formation and their study 
led to conjecture~\cite{SILE} that all the silent universes must be either 
Bianchi I, or Szekeres, or FLRW dust models.  Hence,  it turns out that 
we have proved here this conjecture for the particular case in
which ${}^3 R^a{}_{bcd}=0$.  In other words we have shown that
{\em ``All the silent universes with flat spatial geometry 
belong to the families of Bianchi I,  Szekeres, and FLRW dust models"}.
The only previous partial proof of the conjecture, for the vacuum
case, has been given in~\cite{MARS}.  Therefore, our result is the
first partial proof of the conjecture in the non-vacuum case.

The procedure followed in this work can be applied to other physically
interesting classes of space-times.  By one hand we can
consider a more general energy-momentum content, specifically,
irrotational perfect-fluid models with a geodesic velocity field (in this 
case the pressure depends only on the proper time $\tau$).  Obviously, 
this includes the dust case with a non-zero cosmological constant.  Here,
it would be interesting to investigate whether or not
these models belong to the classes of Bianchi I and Szafron~\cite{SZAF}
space-times (the generalization of the Szekeres models for a perfect-fluid
matter content).  On the other hand,  we can soften the flatness assumption
by demanding the hypersurfaces orthogonal to the fluid velocity to have 
constant curvature, 
i.e., ${}^3 R_{abcd}= K(\tau)(h_{ac}h_{bd}-h_{ad}h_{bc})$, where 
$K(\tau)$ denotes the Gaussian curvature.  Since the models included in 
these two extensions (which can obviously carried out simultaneously)
satisfy (\ref{seco},\ref{divh}), the procedure we have to follow is 
essentially the same.  Another issue that would deserve more attention
is the question of finding a covariant proof of the results presented
here, specially taking into account the fact that these models admit an
intrinsic and covariant characterization.  This would consist in 
pushing forward the study made in section~\ref{cova}, perhaps by
using computer algebra.

Finally, we want to point out that using the covariant space-time
characterization that we have obtained for the 
solutions~(\ref{biai},\ref{cla1},\ref{cla2}), it is possible to find 
a covariant initial-data characterization, which consists in 
considering the local initial-value problem for IDMs and to find the 
initial data whose development corresponds with these models.  However, 
this is beyond the scope of this paper.  An analysis of the initial 
data for some classes of IDMs, including those considered here, will 
be presented in~\cite{SOPU}.

\ack

Some of the calculations in this paper were done using the
computer algebraic system REDUCE.
The author wishes to thank the Alexander von Humboldt Foundation for 
financial support and the Institute for Theoretical Physics of the Jena 
University for hospitality during the first stages of this work.  
Currently, the author is supported by the European Commission 
(contract HPMF-CT-1999-00149).


\end{document}